\documentclass[letter,journal]{IEEEtran}
\usepackage{amsmath,amsfonts}
\usepackage{algorithmic}
\usepackage{algorithm}
\usepackage{array}
\usepackage[caption=false,font=normalsize,labelfont=sf,textfont=sf]{subfig}
\usepackage{textcomp}
\usepackage{color}
\usepackage{stfloats}
\usepackage{url}
\usepackage{verbatim}
\usepackage{graphicx}
\usepackage{cite}
\usepackage{booktabs}
\usepackage{balance}

\begin{document}

\title{Movable Antennas Enabled ISAC Systems: Fundamentals, Opportunities, and Future Directions}

\author{Zhendong Li, Jianle Ba, Zhou Su, Jinyuan Huang, Haixia Peng, Wen Chen, Linkang Du, and Tom H. Luan
\thanks{Z. Li, J. Ba, J. Huang and H. Peng are with the School of Information and Communication Engineering, Xi'an Jiaotong University, Xi'an 710049, China (email: lizhendong@xjtu.edu.cn, 1650376377@stu.xjtu.edu.cn, 2246112066@stu.xjtu.edu.cn, haixia.peng@xjtu.edu.cn). Z. Su, L. Du and T. H. Luan are with the School of Cyber Science and Engineering, Xi'an Jiaotong University, Xi'an 710049, China (email: zhousu@ieee.org, linkangd@xjtu.edu.cn, tom.luan@xjtu.edu.cn). W. Chen is with the Department of Electronic Engineering, Shanghai Jiao Tong University, Shanghai 200240, China (e-mail: wenchen@sjtu.edu.cn). 

(\emph{Corresponding author: Zhou Su})} }

%

\maketitle

\begin{abstract}
The movable antenna (MA)-enabled integrated sensing and communication (ISAC) system attracts widespread attention as an innovative framework. The ISAC system integrates sensing and communication functions, achieving resource sharing across various domains, significantly enhancing communication and sensing performance, and promoting the intelligent interconnection of everything. Meanwhile, MA utilizes the spatial variations of wireless channels by dynamically adjusting the positions of MA elements at the transmitter and receiver to improve the channel and further enhance the performance of the ISAC systems. In this paper, we first outline the fundamental principles of MA and introduce the application scenarios of MA-enabled ISAC systems. Then, we summarize the advantages of MA-enabled ISAC systems in enhancing spectral efficiency, achieving flexible and precise beamforming, and making the signal coverage range adjustable. Besides, a specific case is studied to show the performance gains in terms of transmit power that MA brings to ISAC systems. Finally, we discuss the challenges of MA-enabled ISAC and future research directions, aiming to provide insights for future research on MA-enabled ISAC systems.
\end{abstract}


\section{Introduction}
\IEEEPARstart{T}{he} development of information technology drives the evolution from fifth-generation (5G) to the sixth-generation (6G). It requires that 6G extend its service from connecting people and things to efficiently interconnecting intelligent agents, achieving a transition from the Internet of Things (IoT) to the intelligent IoT. In terms of the communication function, it demands high data rates, low latency, high reliability and support for massive connections. Regarding the sensing function, it requires high-precision positioning, high-resolution imaging, real-time sensing and dynamic monitoring. Integrated sensing and communication (ISAC), as an innovative wireless network paradigm, integrates the functions of communication and sensing. It is capable of providing high-quality wireless communication to target users while offering high-precision sensing services \cite{9737357}. By integrating the communication and sensing functions, ISAC can share spectrum resources and hardware platforms to achieve cost reduction and performance improvement. However, a single-antenna system cannot fully meet the further enhanced requirements for communication and sensing of ISAC in 6G. Besides, it also shows limitations in the efficient utilization of spectrum resources and the adaptability to dynamic environments. 

Multiple-input multiple-output (MIMO)-based ISAC system is proposed as one of the solutions to address the aforementioned limitations. Owing to the deployment of multiple antennas at both the transmitter and receiver, the integration of MIMO with ISAC is expected to simultaneously ensure high resolution and robust sensing performance, high-quality wireless communication \cite{10557534}. Furthermore, by precisely controlling the direction of signal propagation through beamforming, it effectively reduces interference while achieving efficient utilization of spectrum resources, thereby promoting the rapid development of various emerging 6G applications. Under the traditional MIMO architecture, the dynamic adjustment capability of antennas is limited due to their uniform distribution at half-wavelength intervals. To further explore the potential the ISAC systems, antenna selection scheme for traditional fixed antennas is proposed. This scheme primarily involves the selection of one or a group of antennas with optimal performance from multiple fixed positions for signal transmitting and receiving. However, its ability to improve channel conditions is limited, and its dynamic adjustment capability remains insufficient when facing complex and variable communication and sensing environments. Specifically, in terms of spatial resource utilization, it is challenging to flexibly adjust the working mode of the antenna array based on factors such as the distribution of targets in actual scenarios and signal propagation characteristics. Moreover, it means the potential for spatial resource utilization is not fully exploited.

To more fully exploit spatial resources, the movable antenna (MA) is introduced as a novel paradigm \cite{10839251}. In contrast to traditional MIMO systems, the MA benefits from the mobility of antenna elements, which allows for a more effective utilization of the spatial degrees of freedom (DoF) in the transmitter and receiver to dynamically adjust the wireless channel. It is worth noting that moving MA results in non-negligible delay, which may decrease the effective data transmission time \cite{li2024minimizingmovementdelaymovable}. By selecting the optimal positions for antenna elements, the system performance can be further enhanced. For instance, \cite{guo2024movableantennaenhancednetworked} exhibited the potential of MA in enhancing the performance of ISAC systems. It constructed an MA-enhanced networked full-duplex ISAC system model. By jointly optimizing parameters such as beamforming and power allocation to design algorithms, it proved that the MA could change the signal propagation path and phase by flexibly adjusting its position. Meanwhile, with the advent of massive MIMO, the number of antennas and radio frequency (RF) chains required by traditional MIMO is significantly increased. MA is expected to utilize the characteristic of flexible movement of antenna elements to achieve the same performance with fewer antenna elements \cite{10318061}. Therefore, with the advantages of MA, MA-enabled ISAC systems exhibit substantial spatial DoF and inherent sensing capabilities.

Although MA-enabled ISAC systems demonstrate promising advantages and potential performance enhancements, research on MA-enabled ISAC systems is still at the initial stage. This paper focuses on providing an overview of the fundamentals, advantages, challenges and future research directions of MA-enabled ISAC systems. In Section II, we first expound the basic principles of MA, including the MA model and hardware architecture, then classify MA based on the movable spatial dimensions of MA elements and compare their performance with traditional fixed antennas, ultimately discussing the application scenarios of MA-enabled ISAC systems. Then, We analyze the advantages of MA-enabled ISAC systems in Section III. In Section IV, the performance differences between the MA-enabled ISAC systems and the traditional fixed antenna ISAC systems are compared. Simultaneously a discrete antenna positioning algorithm for MA elements is proposed, which is founded on the binary particle swarm optimization (BPSO) framework. The challenges and potential research directions are outlined by us in Section V. Finally, in Section VI, we summarizes this paper.

\section{Overview of MA}
In this section, we mainly introduce the basic model and hardware structure of MA, types of MA, comparison between MA and traditional fixed antennas, as well as the application scenarios where MA enables ISAC systems.  

\subsection{The Basic Models and Hardware Architectures of MA}

\begin{figure}
    \centering
    \includegraphics[width=9cm]{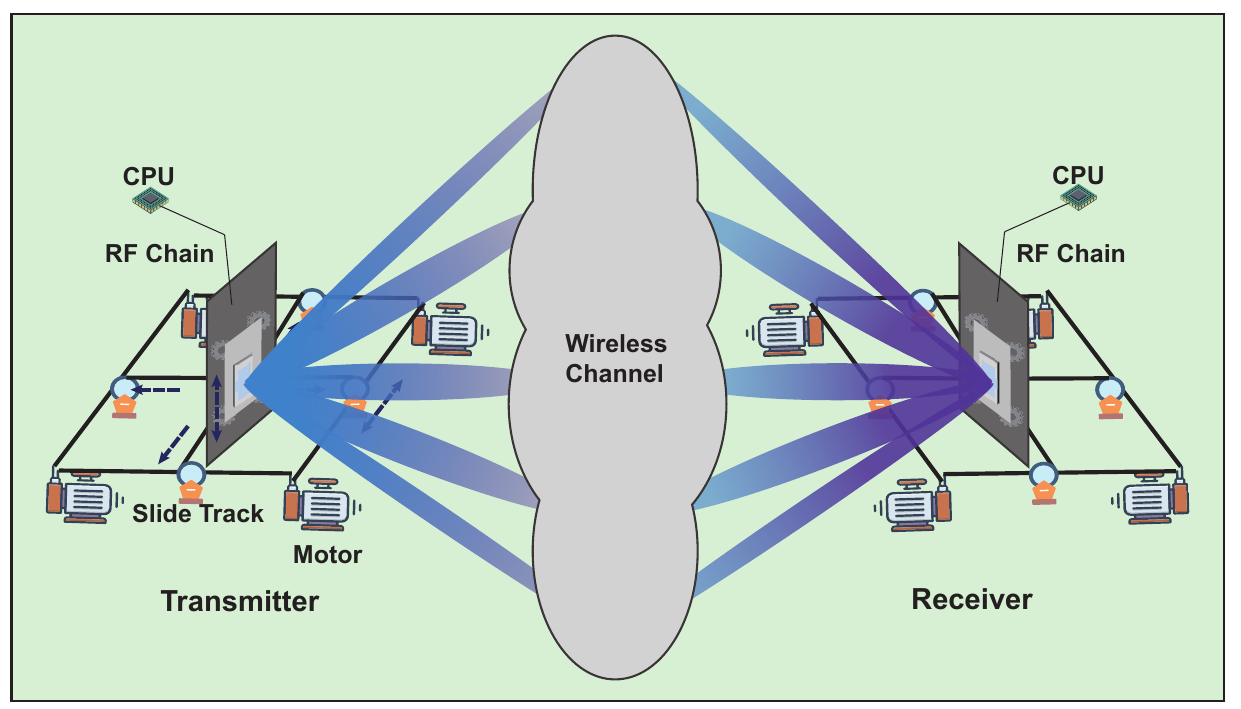}
    \caption{Basic models and hardware architectures of MA.}
\end{figure}

In this subsection, we primarily introduce the MA model and its hardware structure. As shown in Fig. 1, MA is a wireless system where the positions of antenna elements can be flexibly adjusted within a certain range. Compared to traditional fixed antennas, the flexibility of its antenna elements enhances the system's spatial DoF, allowing it to adjust channel conditions, thereby potentially improving signal intensity and reducing attenuation. The movement of MA elements is mainly achieved through micro-electromechanical systems (MEMS) and stepping motors. Early MA was endowed with motion or rotation capabilities through MEMS technology, and the stepping motor-driven MA developed in 2015 can flexibly adjust the antenna positions in radar systems. Then, taking the transmitter and receiver driven by a stepper motor in a communication system as an example, the specific driving process is shown in Fig. 1. Its working principle is consistent with that of traditional fixed antennas, mainly working based on the principle of electromagnetic induction. The driving system is also similar to that of traditional fixed antennas, mainly realizing the connection between MA elements and RF chains through a flexible cable. Theoretically, MA elements are located on a slider that can move in a three dimensional (3D) space. When the central processing unit (CPU) receives communication signals, it processes the signals and issues control instructions to realize the movement of MA elements in the 3D space. Different movable spatial dimensions of MA elements result in different spatial DoF. MA can be deployed in base station, radar receiver and user equipment to achieve high-precision sensing, adaptive beamforming, and better interference management.

\subsection{Types of MA}
Based on the differences in the dimensions of the moving space of MA elements, we classify the MA in this subsection.

\textit{One Dimensional MA (1DMA):} Fig. 2a) comprises $N$ MA elements, with the mobility space of the MA elements being a one dimensional (1D) line segment of length $L$. \cite{10643473} optimized the positions of the 1DMA elements to enhance wireless sensing performance, derived the CRB for angle estimation in 1D  arrays. Furthermore, global optimal solutions and alternating optimization algorithms were developed to optimize antenna positions, thereby significantly improving the accuracy of angle estimation. Practically considered, the movement of 1DMA elements is relatively straightforward and easy to implement. However, the DoF for adjustment is relatively limited, and it is mainly applicable to scenarios where the beam needs to be adjusted in a single direction.

\textit{Two Dimensional MA (2DMA):} As shown in Fig. 2b), it contains $N$ MA elements, and the movable space of MA elements is rectangular. The MA elements can move in both the horizontal and vertical directions. A MA base station architecture was proposed in  \cite{10741192}. By jointly optimizing the positions of 2DMA, the receiving power and the transmitting power, the minimum achievable rate of the multi-user communication system was effectively improved. Generally speaking, compared with 1DMA, it has stronger mobility flexibility, greater spatial gain and can form more directive beams.

\begin{figure}
    \centering
    \includegraphics[width=9cm]{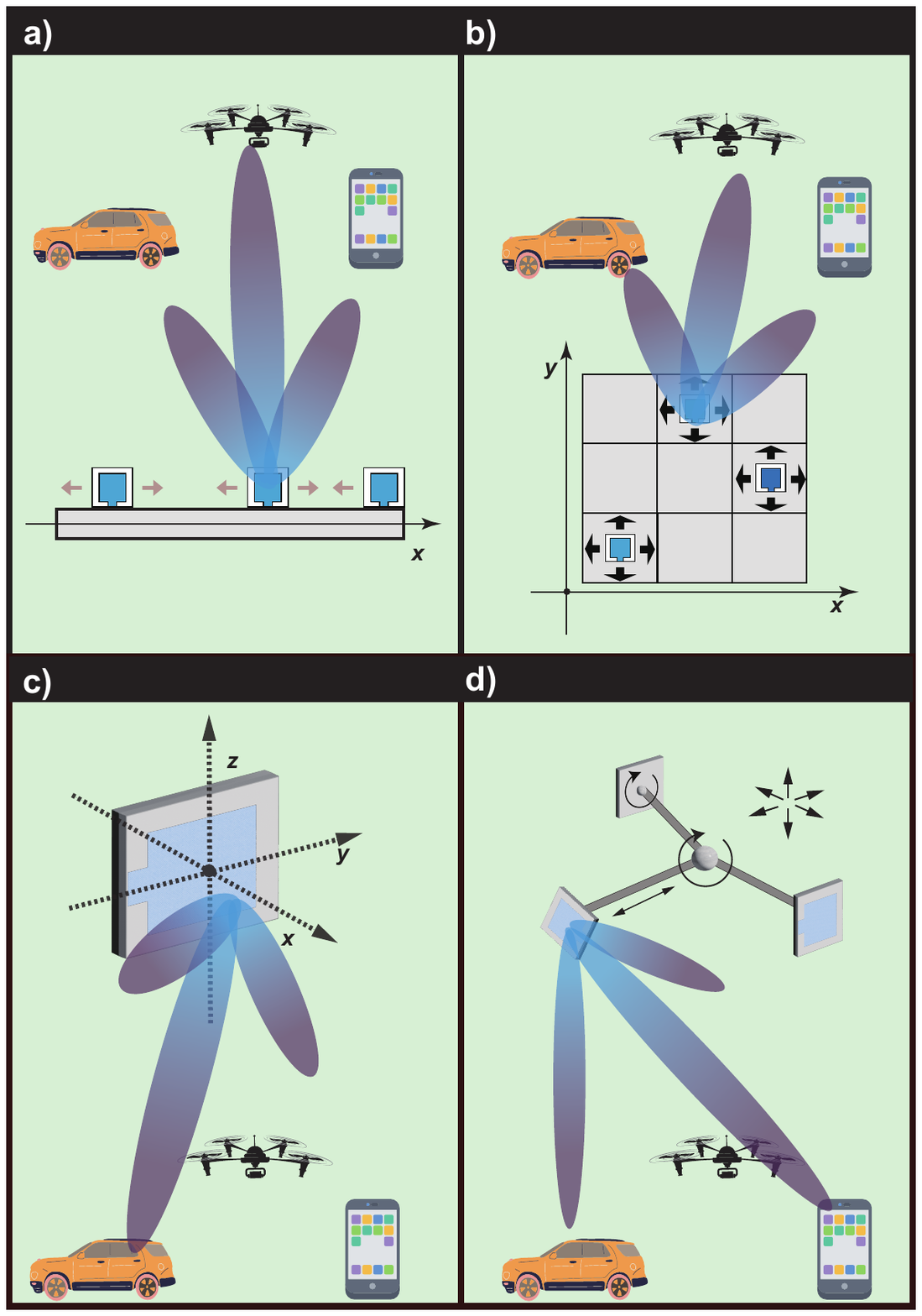}
    \caption{Types of MA: a) 1DMA; b) 2DMA; c) 3DMA; d) 6DMA.}
\end{figure}

\textit{Three Dimensional MA (3DMA):} There are $N$ MA elements and the movable space of MA elements is a cuboid as is illustrated in Fig. 2c). The MA elements can move not only in the directions of the $x$-axis and $y$-axis but also along the $z$-axis. \cite{10354003} studied the application of 3DMA technology in multi-user communication. By optimizing the positions of antennas, they decreased the total transmitting power of users and enhanced the system performance via algorithms. Compared to the aforementioned MA configurations, the increased DoF enable more precise beam steering, thereby achieving higher-performance communication and sensing.

\textit{Six Dimensional MA (6DMA):} It can be observed from Fig. 2d) that, in contrast to the previously mentioned MA configurations, the 6DMA utilizes a uniform linear array or a uniform planar array for the antenna panel rather than mobile MA elements. The MA panel not only achieves 3D movement but also enables 3D rotation through the retraction of rods and the rotation of motors at both ends of the rods, thus expanding the mobility space of the MA to three dimensions. The capacity of wireless networks could be maximized by independently adjusting the 3D position and rotation of antennas using 6DMA, as \cite{10752873} demonstrated. And this led to a significant enhancement in network performance compared to traditional fixed antennas. Due to the further increase in its spatial DoF, greater beam gain and more flexible beamforming can be obtained.

In summary, it can be observed that as the DoF of antenna mobility increase, the resulting spatial gain is greater, but the associated hardware costs also differ. The selection of MA types should be determined by considering specific requirements, costs and multiple other factors.
\newcommand{\tabincell}[2]{\begin{tabular}{@{}#1@{}}#2\end{tabular}}
\begin{table}
	\begin{center}
		\renewcommand{\arraystretch}{}
		\caption{Comparison between 
            MA and Traditional Fixed Antennas}
		\label{T1}
		{\begin{tabular*} 
            {1.005\linewidth}
            {|c|c|c|c|} 
            \hline
            \tabincell{c}{Type of Antenna}&\tabincell{c}{Traditional\\ Fixed Antennas} &\tabincell{c}{(1D-3D)MA} &\tabincell{c}{6DMA}  \\
            \hline
             \tabincell{c}{Mobility of Antenna}&\tabincell{c}{Immovable} &\tabincell{c}{Movable} &\tabincell{c}{Movable}  \\
            \hline
             \tabincell{c}{Mobility of\\ Antenna Elements}&\tabincell{c}{Immovable} &\tabincell{c}{Movable} &\tabincell{c}{Immovable}  \\
            \hline
             \tabincell{c}{Capability of\\ Adjusting Channels}&\tabincell{c}{Weak} &\tabincell{c}{Strong} &\tabincell{c}{Strong}  \\
            \hline
             \tabincell{c}{Capability of\\ Interference Resistance}&\tabincell{c}{Weak} &\tabincell{c}{Strong} &\tabincell{c}{Strong}  \\
            \hline
             \tabincell{c}{Coverage Range}&\tabincell{c}{Non-adjustable} &\tabincell{c}{Adjustable} &\tabincell{c}{Adjustable}  \\
            \hline
             \tabincell{c}{Performance Gain}&\tabincell{c}{Low} &\tabincell{c}{High} &\tabincell{c}{High}  \\
            \hline
            \tabincell{c}{Difficulty of \\Channel Estimation}&\tabincell{c}{Low} &\tabincell{c}{High} &\tabincell{c}{High}  \\
            \hline
            \tabincell{c}{Hardware Cost}&\tabincell{c}{Low} &\tabincell{c}{High} &\tabincell{c}{High}  \\
            \hline
		\end{tabular*}}
	\end{center}
\end{table}

\subsection{Comparison between MA and Traditional Fixed Antennas}
We mainly conduct a comparison between traditional fixed antennas and different types of MA, as shown in Table I. Based on the characteristics of the MA mentioned above, it breaks free from the limitation of the fixed position of traditional fixed antennas, and the spatial DoF are increased. The 1D-3DMA can dynamically adjust the channel by moving the MA elements to improve the performance of the system. Although the 6DMA cannot move the positions of the MA elements, it can dynamically adjust the channel through the movement of the antenna panel. In terms of coverage range, due to its fixed position, the traditional fixed antennas have a relatively limited and fixed coverage range. For MA, whether it is 1D-3DMA or 6DMA, because of their mobility, they can change their positions and postures within a certain space. As a result, the coverage range can be dynamically adjusted according to specific application scenarios and signal requirements. From the perspective of performance gain, for 1D-3DMA and 6DMA, they can utilize the movement of the antenna to concentrate the gain in the target direction, improve the receiving and transmitting intensities of signals, and thus obtain better communication effects. However, the channel state information (CSI) of MA needs to cover the entire spatial area that may be accessed. The larger the movable space of the MA elements is, the greater the difficulty of channel estimation will be. In addition, the movement of the MA elements requires hardware to drive, and its hardware cost is higher under the condition of the same number of antennas and so on. Nevertheless, the proposal of MA still brings new opportunities to the field of wireless communications, and the integration with ISAC and other technologies will bring broader application values. 

\subsection{Application Scenarios of MA}

\begin{figure*}
    \centering
    \includegraphics[width=18.3cm]{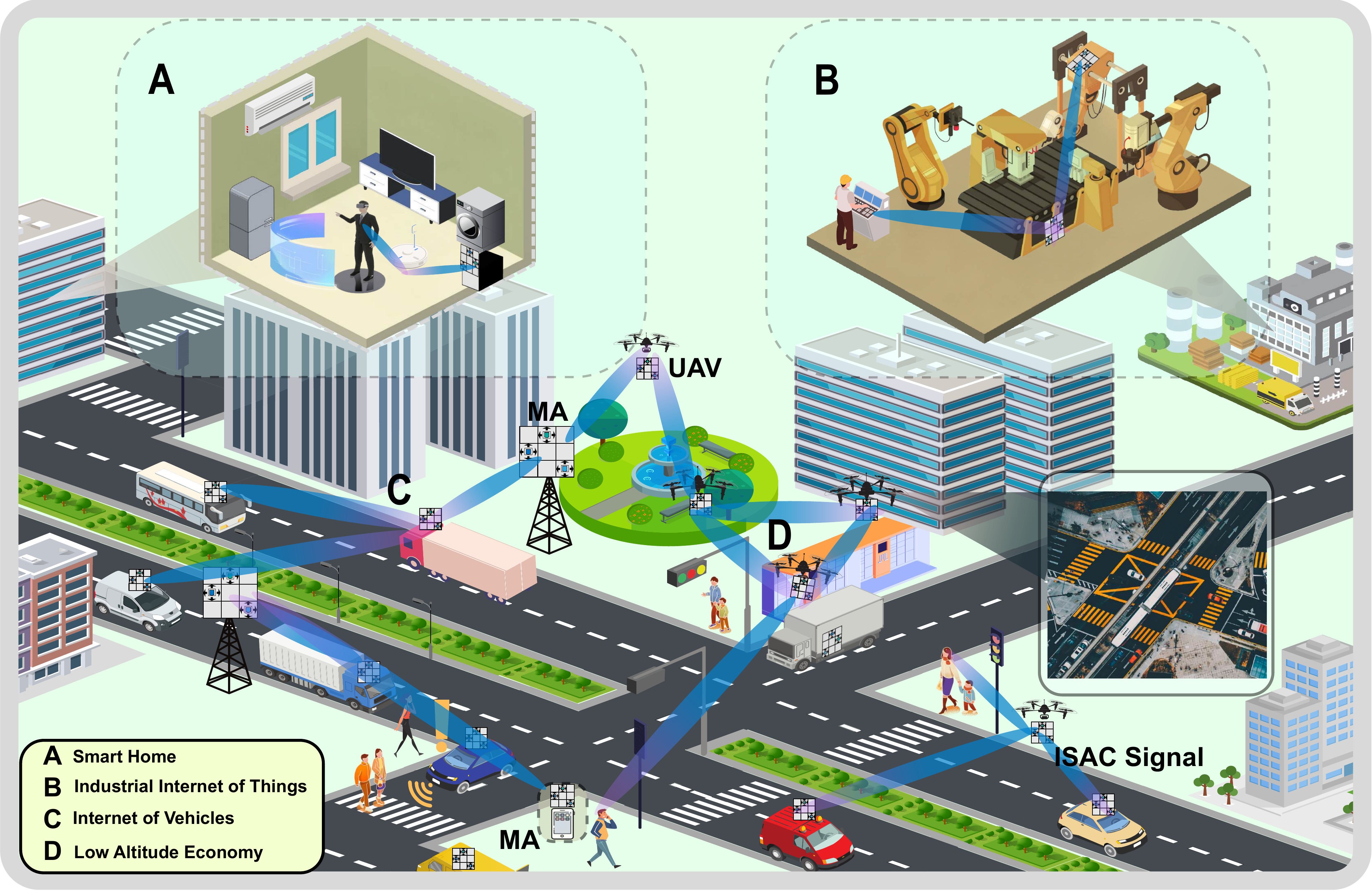}
    \caption{Application scenarios of MA-enabled ISAC systems.}
\end{figure*}

Fig. 3 illustrates the typical application scenarios of MA-enabled ISAC systems, mainly including the following aspects:

\textit{Smart Home:} Taking the residence as a platform, it integrates various facilities in the living environment and realizes automation and remote control through communication among different devices \cite{10256068}. The MA-enabled ISAC system can enhance the interactivity between people and devices. It can locate indoor devices with high-precision sensing capabilities. Meanwhile, MA can optimize signal coverage by dynamically adjusting the positions of antenna elements, and it can maintain stable communication and sensing performance even in complex indoor environments. Ultimately, it helps to improve the security, convenience and comfort of the user's living environment.

\textit{Industrial Internet of Things (IIoT): }This primarily involves achieving intelligent production through communication between equipment and sensors within factories \cite{9526760}. The application of MA and ISAC can take advantage of the ISAC as well as the adjustability of channels to improve signal quality, especially in scenarios with a large number of metal devices and complex electromagnetic interference. MA is capable of dynamically adjusting its position to avoid signal blockage and reflection. Factories can utilize this advantage to perform efficient collaborative work between various devices and issue timely warnings to ensure production safety.

\textit{Internet of Vehicles (IoV):} It mainly provides vehicles with multi-functional services through communication among vehicles as well as between vehicles and information networks \cite{10171185}. The application of MA-enabled ISAC systems and the IoV is expected to offer new ideas for addressing the above challenges. MA can dynamically adjust the antenna positions according to the relative positions and speeds of vehicles to reduce the multipath effect and signal interference, thereby improving the efficiency and safety of traffic flow. It also plays an important role in creating a safe and convenient traffic environment.

\textit{Low Altitude Economy (LAE):} It is an economic form that relies on low altitude airspace and drives the development of various industries dominated by the general aviation industry \cite{10723207}. Facing post-disaster on-site environments and the like, the MA-enabled ISAC system can utilize unmanned aerial vehicles (UAV) and other means to carry out timely rescue. Specifically, by leveraging the mobility of MA, the UAVs can adjust their antenna positions during flight to maintain the optimal communication links. Meanwhile, they can conduct precise environmental perception and target tracking. The application of MA-enabled ISAC systems and the LAE can promote the development of multiple fields.

\section{Advantages of MA-enabled ISAC Systems}
In this section, we mainly analyze the advantages of MA-enabled ISAC systems, which include: improvement of spectrum efficiency, flexible and precise control of beams, and adjustability of signal coverage range.

\subsection{Improvement of Spectrum Efficiency}
In traditional wireless systems, communication systems mainly focus on achieving efficient and stable transmission of signals to meet various data interaction requirements. Meanwhile the sensing function relies on dedicated systems such as radars to detect and acquire information about the surrounding environment. This functional differentiation leads to the situation that communication and sensing respectively occupy different spectrum segments, making it difficult to fully explore and utilize spectrum resources. Moreover, the communication frequency band and the sensing frequency band may be left idle when there is no task, resulting in implicit waste.

The ISAC system, as an innovative paradigm, constructs an ISAC fusion framework by ingeniously integrating waveform designs in the time domain, frequency domain, and spatial domain as well as advanced signal processing techniques. Under this framework, sensing signals and communication messages can be transmitted in parallel within the same frequency band, realizing the sharing and collaborative utilization of spectrum resources. Meanwhile, compared with traditional fixed antennas, the MA still supports MIMO technology and improves spectrum efficiency by utilizing the parallel transmission of multiple data transmission channels. Moreover, it has the ability of dynamic adjustment. It can dynamically adjust channels based on the joint optimization of the position strategies of the transmitting and receiving MA elements, effectively increasing the spatial capacity. The MA and the ISAC system empower each other and generate a synergistic effect, providing support for multiple aspects of wireless communication and promoting the rapid development of multiple fields. 

\subsection{Flexible and Precise Control of Beams}
Flexible beamforming plays a crucial role in the development of modern communication technologies. With the evolution of technology, communication environments become increasingly complex and variable, such as multi-path effects in urban areas, which require dynamic adjustments to optimize signal propagation. In scenarios with multiple users and services, whether to enhance system capacity or meet the requirements of different services, flexible adaptation of beamforming is indispensable. Additionally, in dynamic communication scenarios, support for mobile users and response to temporary communication demands rely on its rapid tracking and adjustment capabilities.

Traditional fixed antennas, on account of their fixed positions, are restricted in beamforming and thus find it hard to adapt to complex and changeable environments. In contrast, in MA-enabled ISAC systems, the sensing function conducts comprehensive detection and analysis of environmental information. Integrating this information deeply into beamforming strategies allows the beamforming vectors to be finely adjusted based on real-time conditions. More importantly, MA elements can change their position and angle in real-time according to the location of mobile terminals, endowing MA with unique spatial optimization capabilities. MA-enabled ISAC systems can adjust the phase and amplitude to control the shape, directionality, and side lobe suppression of the beam. Based on the distribution of high-rise buildings and the density of users in complex urban communication environments, they can optimize the spatial layout of the beam, reducing signal interference and reflection.  Thus, the MA-enabled ISAC system can perform more flexible and precise beamforming, which helps to improve the quality of communication and sensing in complex environments.

\subsection{Adjustability of Signal Coverage Range}
Communication environments become increasingly complex, with urban areas characterized by high-rise buildings that cause signal obstruction and reflection. These scenarios demand that communication systems dynamically adjust the coverage range of signals to maintain communication stability and reliability. This dynamic adjustment capability is crucial for meeting the growing data demands, improving energy efficiency, reducing power consumption, and ensuring communication quality, especially for high-density user areas. Dynamically adjustable signal coverage become a key direction in the development of wireless communication technologies.

Owing to the mobility of MA elements, it has brought a significant impact on the coverage range. The coverage range of traditional fixed antennas shows relatively static characteristics. Its radiation area is mainly defined by the physical parameters determined when the antennas are installed, including factors such as the antenna gain, transmission power and height. However, MA can actively change its position according to different geographical environments and user distribution conditions, greatly expanding the adaptability of the coverage range. For instance, in areas with complex terrains like mountainous regions, it is difficult for traditional fixed antennas to achieve comprehensive coverage. However, MA can be flexibly deployed according to the topographical features to fill in the signal blind areas, enabling areas with weak or no signals to be effectively covered. When in areas such as those with signal shielding, MA can rely on mobility to reconstruct the coverage area, avoid the obstruction of signal propagation caused by interference sources, and ensure the stable transmission of signals within the effective range. By flexibly adjusting the coverage range, it is possible to accurately adapt to changes in the ISAC system's sensing and communication needs in different scenarios.

\section{Case Study}
Based on the above discussion, this section mainly analyzes the advantages of the MA-enabled ISAC system by comparing the MA-enabled full-duplex ISAC system model with the full-duplex ISAC model of fixed antennas. 


In this case, we mainly analyze the problem of minimizing the total transmit power of the MA-enabled full-duplex ISAC system \cite{li2024joint}. The base station is equipped with a receiving MA and a transmitting MA. It is worth noting that we assume that the candidate positions of the MA elements are in a two dimensional (2D) space and are discrete. The movement of the actual electromechanical equipment that can control the movement of the MA elements currently has a limited precision and is discrete, thus making this MA model more practical. Meanwhile, due to the limited adjustable positions of the discrete positions, the search space is small, which is beneficial to finding a better solution. The discrete positions of the MA elements, the beamforming vectors, the covariance matrix of the sensing signals and the user transmit power, which are the optimization variables in this model, are highly coupled. Therefore, the established problem is non-convex. In contrast to prior studies that have utilized continuous candidate positions for the MA elements, these studies mainly employ particle swarm optimization (PSO) as a methodological approach to ascertain the optimal positions of MA elements. Given the adoption of a more practical discrete set of MA candidate positions in this paper, we propose a BPSO-based optimization algorithm framework. The algorithm is designed to address the complexities associated with discrete candidate positions, thereby offering a more tailored solution to the problem at hand. Specifically, we first use an iterative method to determine the exact positions of the MA elements, and this step involves the solution of the fitness function. In the fitness function, for the parts with non-convex properties and rank-1 characteristics, we respectively adopt the difference-of-convex (DC) programming and the successive convex approximation (SCA) techniques to handle. When the termination condition of the BPSO iteration process is reached, the positions of the MA elements are determined. Subsequently, based on these position information, we can derive the specific beamforming vectors, the covariance matrix of the sensing signals and the user transmit power. To evaluate the effectiveness of the proposed algorithm, we compare the performance with that of the traditional fixed antenna and draw the received and transmitted waveform diagrams based on the solved optimization variables. 


\begin{figure}
    \centering
    \includegraphics[width=8.5cm]{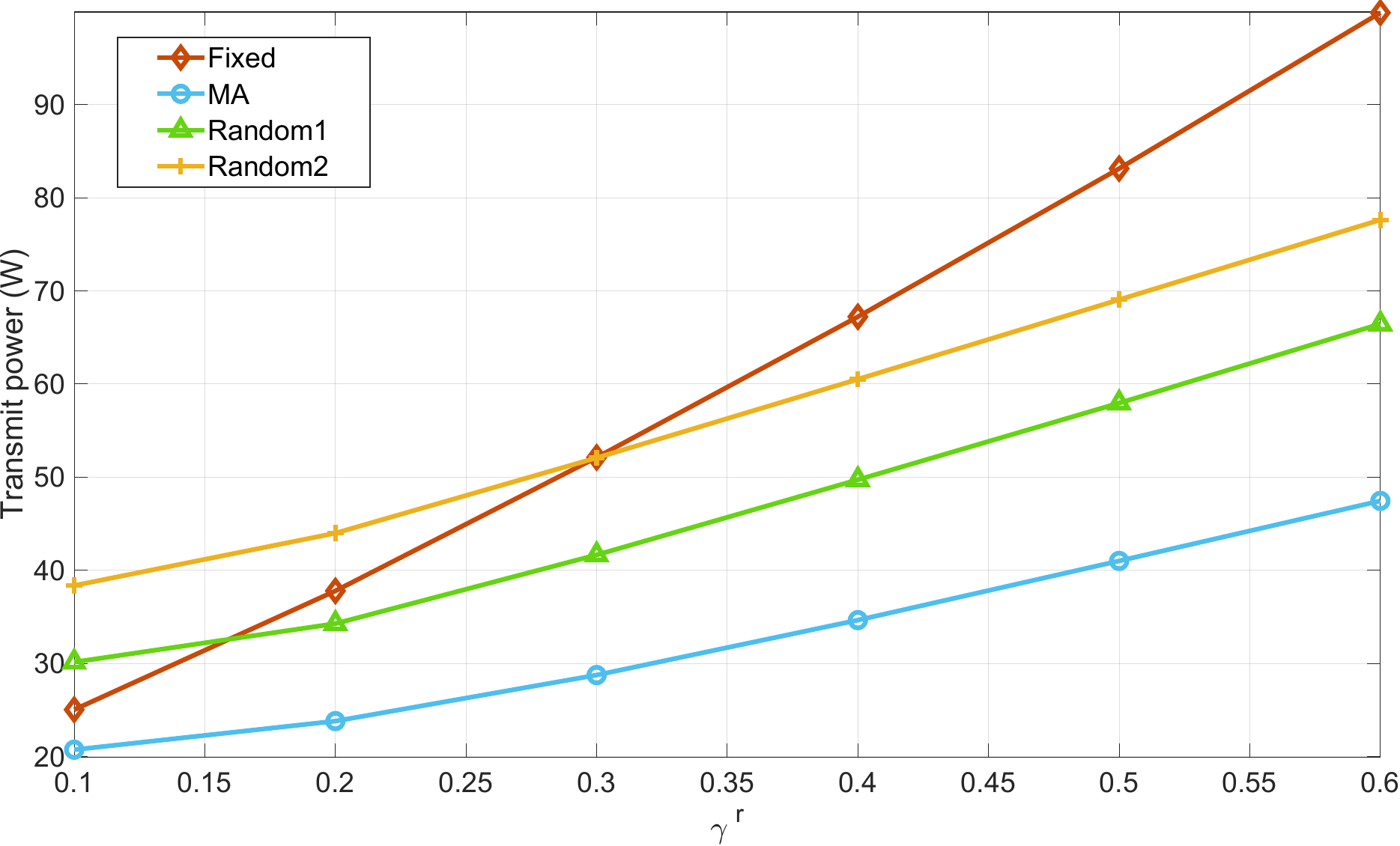}
    \caption{Transmit power consumption versus the sensing SINR threshold.}
\end{figure}

\begin{figure}
    \centering
    \includegraphics[width=9.5cm]{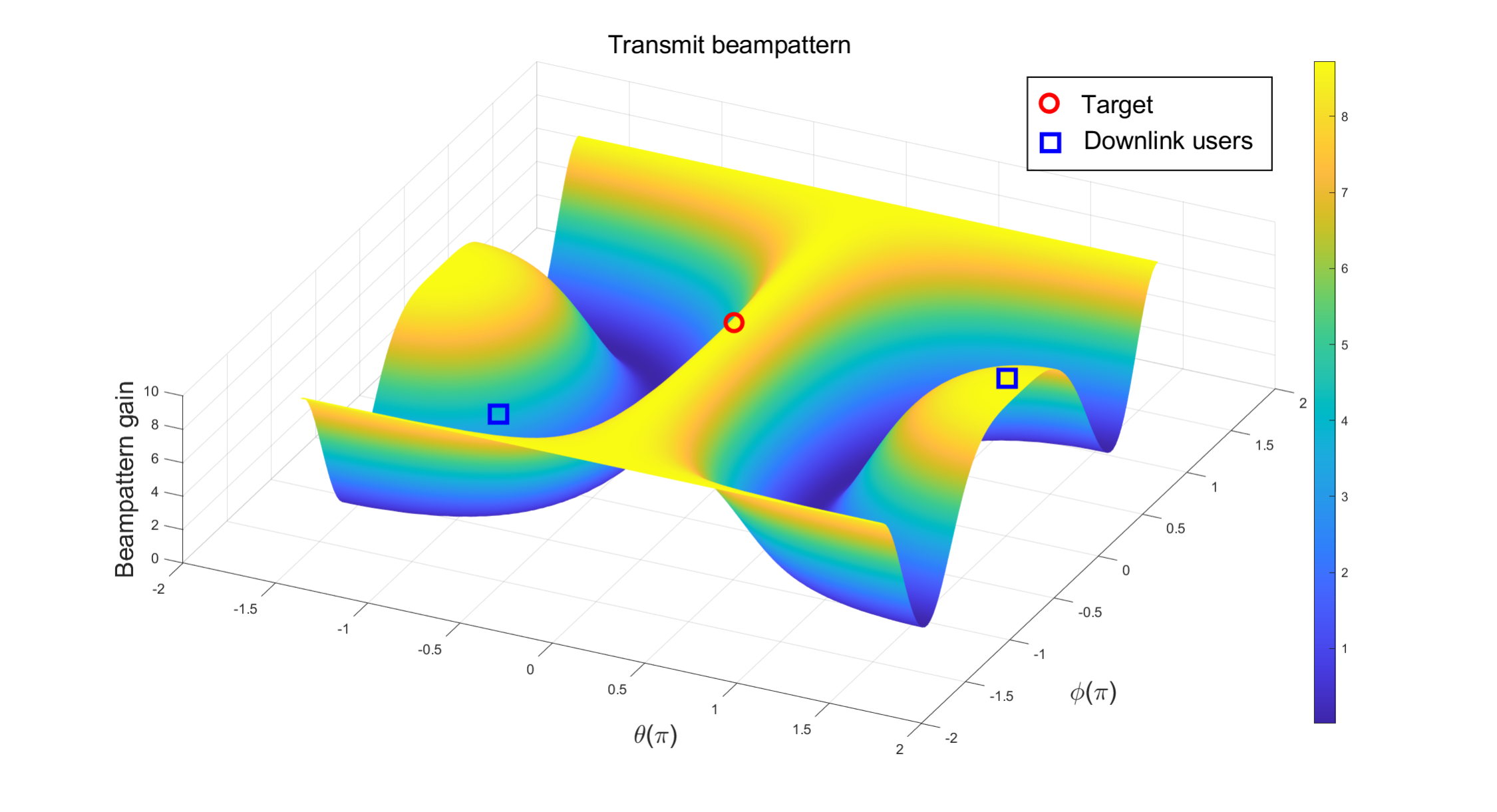}
    \caption{Transmit beampattern regarding radar sensing functionality.}
\end{figure}
During the numerical simulation, we assume that the candidate discrete positions of the MA elements are distributed on a 2D plane and the number of these positions is 9, while the number of MA elements is 2. Meanwhile, the number of both downlink users and uplink users is 2, the number of target sensing radars is 1, and the number of non-target sensing radars is 2. Fig. 4 depicts the relationships between the sensing signal-to-interference-plus-noise ratio (SINR) constraints and the transmit power. Random1 and Random2 are antenna arrays with two randomly selected candidate discrete positions, and Fixed represents the traditional fixed antenna array, and MA stands for the MA array. It can be seen from the Fig. 4 that, due to the advantage that MA can improve the spatial DoF and dynamically adjust the channels, the MA-enabled ISAC system generally has a lower total transmit power than the other three under different sensing constraints. Fig. 5 is the waveform diagram of the transmitted signals drawn based on the optimization variables obtained from the above algorithm. It can be seen from the Fig. 5 that the main transmitting beams are respectively directed towards the target users and the downlink users. In summary, since MA increases the spatial DoF of the system, compared with the traditional ISAC system, the MA-enabled system can reduce the total transmit power more effectively. Moreover, the MA-enabled ISAC system can achieve beam alignment and a certain degree of interference suppression.

\section{Challenges and Future Research
Directions}

In this section, we propose the main challenges of MA-enabled ISAC systems in terms of evaluation criteria, MA positioning, channel estimation, as well as security and privacy, and also proposed potential research directions. 

\subsection{Evaluation Criteria}
Traditional performance evaluation criteria only cover aspects like the data transmission efficiency, spectrum efficiency, bit error rate of communication function, and sensing accuracy and resolution of sensing function. However, these indicators are insufficient for evaluating the MA-enabled ISAC system. Compared with traditional systems, it requires additional measurement indicators. Due to MA's mobility, for antenna performance, as the gain changes with different positions and postures, the dynamic range of antenna gain must be considered. For sensing performance, the sensing range of the MA system changes dynamically with the antenna's position during movement, so the change of the dynamic sensing range should be noted. Similarly, for communication performance, it involves the handover success rate and delay, etc. Also, considering MA's mobility, multi-dimensional factors like its adjustment time and power consumption must be comprehensively taken into account to achieve a more accurate performance evaluation for MA-enabled ISAC systems to diverse functional requirements. Meanwhile, when designing the MA array, the feasible movement area and the minimum spatial interval need to be carefully considered. They respectively define the range within which the antenna elements can move and affect the characteristics of the sidelobe level, thereby affecting the dynamics and accuracy of beamforming.

\subsection{MA Positioning}
Compared with traditional fixed antennas, the MA has a unique feature in that its elements are movable. Owing to this movability, the MA can enhance the spatial DoF and thus improve the performance of the system. However, determining the positions of MA elements is a crucial issue. The optimization of MA elements positions needs to take into account both communication targets and sensing targets. Since variables are generally coupled, such optimization problems often exhibit non-convex characteristics. Under the condition of ideal CSI, the most traditional approach can be adopted, that is, performing alternating optimization by fixing the positions of other MA elements and only moving the position of one MA element. When there are a large number of MA elements and the situation is rather complex, traditional algorithms may be time-consuming. The particle swarm optimization (PSO) algorithm, machine learning are expected to provide ideas for this challenge. Meanwhile, if the candidate positions of MA are continuous, it means that there are theoretically an infinite number of possible position choices, which poses huge challenges to both the implementation of optimization algorithms and the consumption of computing resources. How to improve the efficiency of optimization algorithms through discretization methods is of great importance. 

\subsection{Channel Estimation}

Channel estimation plays a crucial role in wireless communication. It provides channel information for techniques such as beamforming, enabling flexible resource allocation and effectively alleviating adverse impacts like channel fading, thereby improving the performance of wireless systems. However, in MA-enabled ISAC systems, due to the need to simultaneously meet the dual requirements of communication and sensing, the movement of MA elements causes the channel to undergo continuous dynamic changes, which significantly increases the complexity and difficulty of channel estimation, especially when the moving space of MA elements is extensive. Traditional methods such as the least squares estimation have problems in that it has poor adaptability to dynamic changes and is easily affected by noise. The pilot-based channel estimation is troubled by the mismatch between pilot resources and the dynamic channel as well as the high complexity of pilot design. The compression sensing-based channel estimation faces issues such as the destruction of the sparsity assumption and the influence of dynamic changes on the recovery algorithm, making them difficult to be applied to this system. Consequently, conducting effective channel estimation for MA-enabled ISAC systems is a research direction of great value. 

\subsection{Security and Privacy}
Since the MA-enabled ISAC system integrates the two core functions of sensing and communication, the signals it transmits include information for both sensing and communication. Moreover, it needs to receive radar echoes carrying target sensing information, which gives rise to security and privacy issues. Attackers can directly launch attacks on the transmitted signals and the echo signals, causing the communication and sensing functions between the transmitter and the receiver to fail to operate normally. Meanwhile, compared with traditional communication systems, when eavesdroppers intercept the ISAC system, they can obtain communication data. In addition, they can also obtain sensing information about targets or the surrounding environment. Such information is sensitive, and once leaked, it is highly likely to lead to serious consequences such as privacy leakage, posing a major threat to the system security and user privacy. The MA-enabled ISAC system is expected to build a more robust security protection mechanism.
It can achieve this by jointly optimizing the positions of MA elements, beamforming vectors and adopting encryption means, so as to effectively enhance the overall security of the system and resist potential attack and eavesdropping risks.

\section{Conclusion}
This paper has outlined a framework of the MA-enabled ISAC system. Compared with communication systems and sensing systems equipped with traditional fixed antennas, due to the mobility of MA and the integration of communication and sensing in ISAC, its spectrum efficiency is improved, beamforming is more precise and flexible, and the signal coverage range is adjustable. Moreover, we have verified the performance gain in transmit power consumption that MA brings to the ISAC system through a specific case study. Based on the above advantages, it is more capable of adapting to the complex and changeable environment and meeting the increasingly high communication and sensing requirements. Since that the research on MA-enabled ISAC systems is still in its infancy and needs to be further explored and improved in many aspects, it is hoped that this paper can provide valuable references for future theoretical research and specific practices. 
\balance
\bibliographystyle{IEEEtran}

\bibliography{reference}

\end{document}